\documentclass[english,aps,pra,showpacs]{revtex4}
\usepackage{graphicx}
\usepackage{babel}

\begin{document}

\markboth{Regnault and Jolicoeur}{Quantum Hall fractions in Atomic
vapors}

%
%

\title{QUANTUM HALL FRACTIONS IN ULTRACOLD ATOMIC VAPORS}

\author{N.~Regnault}

\author{Th.~Jolic\oe ur}

 \affiliation{Laboratoire Pierre Aigrain,
D\'epartement de Physique,\\ 24, rue Lhomond, 75005 Paris,
France\\Nicolas.Regnault@lpa.ens.fr, Thierry.Jolicoeur@lpa.ens.fr}


\begin{abstract}
Atomic vapors can be prepared and manipulated at very
low densities and temperatures. When they are rotating, they can reach a
quantum Hall regime in which there should be manifestations
of the fractional quantum Hall effect. We discuss the appearance
of the principal sequence of fractions $\nu =p/(p\pm 1)$ for
bosonic atoms. The termination
point of this series is the paired Moore-Read Pfaffian state.
Exotic states fill the gap between the paired state and the vortex lattice expected
at high filling of the lowest Landau level.
In fermionic vapors, the $p$-wave scattering typical of
ultralow energy collisions leads to the hard-core model
when restricted to the lowest Landau level.
\end{abstract}

\maketitle


\section{Introduction}

It is now feasible to study atomic vapors, fermionic as well as
bosonic, in magneto-optical traps at very low temperature and in a
dilute regime. Progress in the trapping and cooling of atoms has
led to the observation of the Bose condensation of many elements,
notably all alkalis. This regime can be studied with exquisite
precision since it can be treated by the mean-field theory for
bosons, i.e. the Gross-Pitaevskii equation, something which was not
available for the dense,
strongly interacting, liquid Helium-4. Similarly fermionic vapors
can be cooled down to the quantum degenerate limit where the Fermi
sea becomes dominant. In the Bose case, the cooling can
 be performed efficiently for a single hyperfine species. In the case
of fermions, when there is a single hyperfine species, the ultralow
energy scattering takes place in the partial wave $L=1$ leading to
much weaker interactions. As a consequence, it is more difficult
to cool down fermions. One escape is the sympathetic cooling
involving two hyperfine species.

The cold atoms display a wealth of interesting phenomena when the
trap is rotating~\cite{Matthews99,Madison00,Abo01,Haljan01}.
In the Bose case, the condensed atoms are in a
superfluid state and as such, they have a very special response to
rotation. If the rotation is slow enough the flow is irrotational
but beyond a critical value of the rotation velocity there is
nucleation of a vortex in which essentially all atoms occupy a
state with unit angular momentum along the axis of rotation.
Beyond this regime one creates more vortices and their mutual
repulsion leads to the formation of the celebrated Abrikosov
triangular lattice. This physics takes place when the angular
momentum (in units of $\hbar$) is of the order of the number of
particles. For even faster rotations, it has been remarked that
one should enter a quantum Hall regime~\cite{Wilkin98}. This takes place when the
angular momentum is of the order of the \textit{square} of the number of
particles. Such a regime is reached when the centrifugal force
exactly compensates the harmonic trapping force confining the
atoms and then one is left only with the Coriolis force which is
formally analogous to a magnetic field. If the confinement
\textit{along} the rotation axis is strong enough, then the gas is in
a two-dimensional (2D) regime and this is the arena of the
fractional quantum Hall effect (FQHE). Present experiments have already
reached the required fast rotation limit but it remains to create
a truly 2D regime and to achieve temperatures low enough to
concentrate all atoms in the lowest Landau level (LLL). The huge
degeneracy of the Landau level problem is then lifted only by
interactions, the s-wave collisions in the case of bosons.

Rotation is also an interesting probe of the physics of fermions.
Notably it has been suggested that it may reveal the BCS
condensation expected for attractive interactions. As in the case
of bosons, the condensed state is a superfluid with reduced
momentum of inertia and in the irrotational flow regime there are
signatures of superfluidity in the quadrupolar modes of
oscillations of the atomic cloud. These signatures have been
observed in the case of the Bose condensates~\cite{Cozzini03}.

In the fast rotation regime, fermions also will reach the quantum
Hall regime. The main novelty with respect to condensed matter
physics is now the nature of interactions between the
atoms~\cite{Regnault04L}. When
there is a single hyperfine species
then collisions are p-wave in the low-energy limit~: colliding pairs of
atoms have
relative angular momentum $L=1$. When projected in the lowest landau
level, this is exactly the so-called hard-core model~\cite{Trugman85},
well known
from the very beginning of the studies of the FQHE. For this
interaction, the celebrated Laughlin wavefunction for fermions is
the exact ground state of the system for a special value of the
angular momentum. Many (if not all) of the features of the FQHE in
solid-state devices should thus be accessible to experiments in
this regime. Notably there should be an emergent Fermi sea for
half-filling of the LLL as a termination point of the FQHE series
of fractions $\nu =p/(2p\pm 1)$ to be contrasted with the gapped
paired state of bosons at $\nu =1$.

In section 2, we discuss the basics of the rotating frame
formalism. In section 3, we explain the relevant parameters of
atomic collisions. Section 4 is devoted to the study of FQHE
fractions for bosons. Section 5 discusses the special case of
complete filling of the lowest Landau level by the bosons. Section
6 discusses the transition towards the vortex lattice. Section 7
is devoted to the case of trapped fermions and their FQHE physics.
Finally section 8 contains our conclusions.

\section{Trapped atoms in rotation}

Transformation to the rotating frame can be done by the
substitution $\mathcal{H}\rightarrow\mathcal{H}-\omega L_z $ where
$L_z$ is the total angular momentum along the rotation axis. In
the rotating frame \cite{Rosenbusch02}, the Hamiltonian describing
N trapped atoms of mass $m$ is given by~:
\begin{eqnarray}\label{Ham1}
 \nonumber
  \mathcal{H} &=& \sum^{N}_{i=1}\frac{1}{2m}(\mathbf{p}_{i}-m\omega
    \mathbf{\hat{z}}\times \mathbf{r}_{i})^{2}
    +\frac{1}{2}m(\omega_{0}^{2}-\omega^{2})(x_{i}^{2}+y_{i}^{2}) \\
   &+&\frac{1}{2}m\omega_{z}^{2}z_{i}^{2}
   +\sum_{i<j}^{N}V(\mathbf{r}_{i}-\mathbf{r}_{j}).
\end{eqnarray}
Here we have assumed a harmonic trapping potential in the $xy$
plane with $xy$ trap frequency $\omega_{0}$. There is also
a trapping potential along the z-axis with frequency $\omega_{z}$
and the angular velocity vector is $\omega\mathbf{\hat{z}}$. For
$\omega$ close to $\omega_0$, the physics is that of charge-$e$
particles in a magnetic field
$\mathbf{B}=(2m\omega/e)\mathbf{\hat{z}}$, corresponding to a
magnetic length $\ell=\sqrt{\hbar/(2m\omega)}$. The cyclotron
frequency of the equivalent magnetic problem is thus
$\omega_c=2\omega$. Typically, the trap frequencies are in the
range 10-100 Hz. Condensates may be prepared in very anisotropic potentials,
one or two-dimensional.
The quantum Hall
states will apppear only in an effectively two-dimensional set-up.
This may be achieved by use of an optical lattice along the
$z$-axis. For the typical traps, when rotated at the critical
frequency, the Landau level spacing $\hbar\omega_c$ is of the
order of a few nanoKelvins. Again the quantum Hall phases will require
temperatures below this value to exist.

Present studies of the vortex lattice that forms at low rotation
frequencies use an additional potential which is anisotropic in
the $xy$ plane to stir the condensate up to the required angular
momentum. Once the required regime is reached one can then remove
the stirring potential, going back to full rotational symmetry,
and hope that thermodynamic equilibrium is reached in the rotating
frame after some relaxation time. To go beyond the critical value
for the rotation frequency, one has to add an extra confining
potential to prevent explosion of the gas, e.g. a quartic term is
feasible. It is important to note that one has in general
knowledge neither of the rotation frequency of the gas nor of its
angular momentum. These crucial quantities have to be measured
directly or inferred from related quantities.

\section{Collisions and pseudopotentials}

In the context of ultracold trapped gases, it is important to note
that the collisions take place with a characteristic energy scale
which is set by the temperature, nanoKelvins or less, and many
orders of magnitude less than the energy scales of the internal
degrees of freedom of the interatomic potential. In this limit,
the scattering problem simplifies tremendously and is given by
only a few partial waves. For a single species of bosons, the
scattering takes places only for zero relative angular momentum
and the phase shift may be written as $\delta_0 (k)\simeq -ka_s$
for $k\rightarrow0$, where $k$ is the momentum and $a_s$ is the
s-wave scattering length. This limiting case may be studied by use
of the so-called pseudopotential~:
\begin{equation}
\label{swave}
    V_0(\mathbf{r})=(4\pi\hbar^{2}a_s/m)\delta^{(3)}(\mathbf{r}),
\end{equation}
where $m$ is the mass of the bosonic atoms. This pseudopotential
is known to lead to an ill-defined scattering problem in general
and should be regularized in some cases. However it has the virtue
of reproducing correctly the Born approximation of the true
scattering problem. In the case of the fractional quantum Hall
effect, we need only to diagonalize the interaction problem
restricted to the lowest Landau level for many of the important
questions. Since this is essentially first-order perturbation
theory on a massively degenerate level, it is enough to use the
singular unregularized potential Eq.(\ref{swave}).

In the case of a single species of fermions, the antisymmetry of
the wavefunction implies that the scattering now takes place for
relative angular momentum $L=1$. The relevant phase shift is now
$\delta_1 (k)\simeq -k^3 a_p^3/3$ for $k\rightarrow0$, where $a_p$
is the p-wave scattering length. There is also a pseudopotential
in the p-wave case~:
\begin{equation}\label{pwave}
 \hat{V}_1=(12\pi a_p^3/m)\,\,\,
 \hat{\mathbf{p}}\,\,\delta^{(3)}(\mathbf{r})\,\,\hat{\mathbf{p}},
\end{equation}
where $\hat{\mathbf{p}}$ is the tridimensional impulsion operator.
If there are more than one hyperfine species then again s-wave
scattering is allowed and thus dominates because of the stronger
vanishing of the p-wave phase shift. The smallness of p-wave
interactions is generally considered as a nuisance when cooling fermions.
However it has recently been demonstrated~\cite{Regal03} that it is possible to
tune the scattering length by going through a resonance as a
function of an applied external field. This is similar to the
manipulation of the s-wave scattering by use of a Feshbach
resonance. To obtain a sizeable FQHE gap in a Fermi gas, it may
necessary to enhance the interactions by such a resonance.

\section{The principal sequence}

We now discuss the FQHE physics for a single component Bose gas.
If we are in a 2D regime at the critical frequency defined in
Eq.(\ref{Ham1}) with the potential given by the Eq.(\ref{swave}),
then we are exactly in the conditions required for the apparition
of the FQHE for bosons with delta interactions. To make progress,
it is possible to write down the wavefunctions and the
hierarchical constructions invented for fermions and it is also
possible to perform exact diagonalizations on small systems.
Pioneering studies have been performed in
refs.(\cite{Cooper99,Wilkin00,Cooper01}).
We have also followed these two strategies~\cite{Regnault03,Regnault04B}.
In the numerical studies we
used the spherical geometry which has been shown to be very useful
in the case of fermions with Coulomb
interactions~\cite{Haldane85}. Strictly speaking, the experiments
most closely correspond to the geometry of an unbounded disk and
it is thus important to translate our results also in this set-up.
In the symmetric gauge and in the unbounded plane, the eigenstates
of the lowest Landau level are given by~:
\begin{equation}
\label{LLLf} \phi_m(z) =\frac{1}{\sqrt{\ell\sqrt{\pi}}} \,\,\, z^m
\mathrm{e}^{-|z|^2/4\ell^2},
\end{equation}
where we use the complex notation $z=x+iy$ and $m$ is the angular
momentum of the state along $z$. To reside entirely in the lowest
Landau level, an otherwise arbitrary quantum state should be of
the form~:
\begin{equation}
\label{LLLpsi} \Psi(z_i) =f(z_i) \,\,\,
\mathrm{e}^{-\sum_i |z_i|^2/4\ell^2},
\end{equation}
where $f$ is
an analytic function of the $z_i$'s.

\subsection{The $\nu =1/2$ Laughlin state}

The Laughlin wavefunction for bosons is given by~:
\begin{equation}\label{LJ}
    \Psi^{(m=2)}=\prod_{i<j}(z_i -z_j)^2\,\,\,
    \mathrm{e}^{-\sum_i|z_i|^2/4\ell^2}.
\end{equation}
This state is in the LLL and is an exact zero-energy eigenstate of
the delta interaction~\cite{Wilkin98}. It is the lowest total angular momentum
state with these properties. For higher angular momentum, it is
always possible to multiply $\Psi^{(m=2)}$ by an arbitrary
symmetric polynomial in the $z_i$'s to get more zero-energy
states. These states are edge excitations at least for not too
high momentum. The state Eq.(\ref{LJ}) describes a droplet of
incompressible fluid centered at the origin and the mean density
corresponds to a filling fraction $\nu =1/2$ of the LLL. More
generally the filling fraction appears in the relation between the
number of particles and the total angular momentum~: $L_z
=N^2/(2\nu)$ for large $N$ and $L_z$. A sample spectrum for N=5
bosons is displayed in Fig.(1). The ground states are the Laughlin
states and the edge excitations. For larger angular momentum,
there is also the quasihole excitation which is gapless, a peculiarity of the
delta interaction. The Laughlin fluid also support quasielectron
excitations obtained by reducing the angular momentum by N unit
exactly~: this is the first nonzero energy state in Fig.~(1). It
has descendants with same energy due to the center-of-mass motion
(at no extra cost in the LLL\cite{Trugman85}). This quasielectron
has a gap that remains nonzero in the thermodynamic limit. We have
shown that it is given by $\approx 0.09g$ where
$g=\sqrt{32\pi}\hbar\omega_c a_s/\ell_z$ and $\ell_z$ is the confinement
length along the $z$-axis.

\begin{figure}[!htbp]
\begin{center}\includegraphics[width=3.25in,
  keepaspectratio]{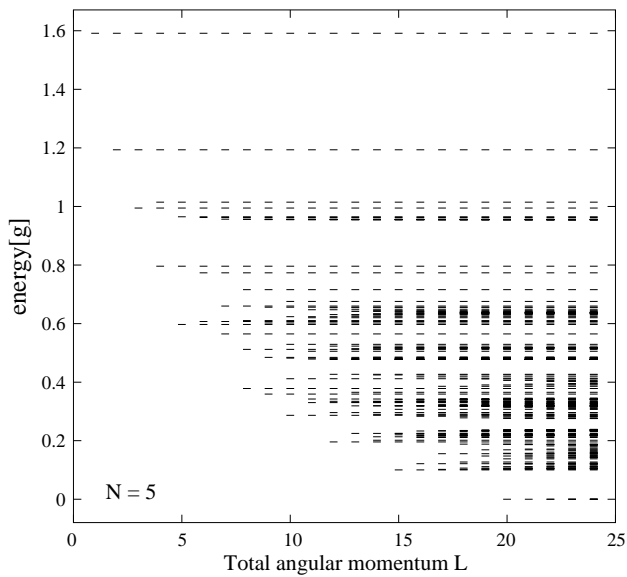}
\end{center}
\caption{Spectrum of N=5 bosons in the unbounded disk geometry.
The Laughlin state is at zero-energy and L$_z$=20.}
\label{disk}
\end{figure}

This nonzero gap also appears in the spectrum of collective
density excitations. This is most clearly seen in the spherical
geometry. The motion of the particles is restricted to a sphere
which experiences uniform flux from its center, as created by a
magnetic monopole. The flux 2S in dimensionless units is quantized
by Dirac's condition~: it is a positive integer. The radius of the
sphere is given by $R=\ell\sqrt{2S}$ where $\ell$ is the magnetic
length. The problem of Landau levels on the sphere was solved long
ago by Tamm~\cite{Tamm31}. The eigenstates are monopole harmonics,
a generalization of spherical
harmonics~\cite{Wu76,Haldane83,Fano86}. The LLL has 2S+1 states,
the well-known Landau degeneracy. Wavefunctions constructed in the
disk geometry may be translated on the sphere by a stereographic
projection~\cite{Fano86}. It is important to note that in general
quantum Hall fluids require a finite shift on the sphere~: the
relation between the flux and the number of particle is of the form
$2S=(N/\nu)-X$ and the quantity $X$ may be computed once we have a
guess for the wavefunction. In the case of the Laughlin
wavefunction, we have $2S=m(N-1)$. Hierarchical constructions give
us definite predictions about the shift that can be then tested by
exact diagonalization. The spherical geometry preserves the full
SU(2) symmetry and thus the eigenstates are classified by their
total angular momentum. Here the Laughlin state is a singlet state
of the \textit{total} angular momentum. It is separated by a clear
gap from a branch of isolated states that abruptly terminates at
$L=N$~: this is what we expect for two-particle states made of
quasiparticles each having angular momentum $L=N/2$.

\subsection{other fractions}
The standard hierarchies may be easily translated to the case of
Bose particles. It is natural to expect that the Bose analog of
the prominent Jain sequence of fractions~\cite{Jain89} should display the most
stable states. This sequence is realized when $\nu =p/(p\pm 1)$.
These states are naturally "explained" in the composite fermion
scheme~: each boson captures a vortex with one quantum of
statistical flux and is transmuted into a fermionic particle, the
composite fermion (CF). In a mean-field picture, the fraction $\nu
=p/(p\pm 1)$ is realized when the CFs occupy exactly $p$ Landau
levels.
 On the sphere we find that there are incompressible
states along the lines given by~:
\begin{equation}\label{Hshift}
    2S=\frac{p\pm 1}{p}N \mp p -1 \quad\mathrm{for}\quad \nu=\frac{p}{p\pm 1}.
\end{equation}
In fig.(2) we have displayed the corresponding candidates.  In the
interval $1>\nu >1/2$, we observe candidate states for $\nu =2/3,
3/4, 4/5$. They have the expected scaling properties towards the
thermodynamic limit\cite{Regnault04B}. These fluids display also a
collective mode which is gapped for all values of the angular
momentum as is the case of the parent $\nu =1/2$ fluid. The
apparition of these fractions follows the standard hierarchical
scheme~: deviation from the commensurability in Eq.(\ref{Hshift})
leads to the creation of extra quasiparticles that finally
condense in a new incompressible fluid.

\begin{figure}[!htbp]
\begin{center}\includegraphics[width=3.25in,
  keepaspectratio]{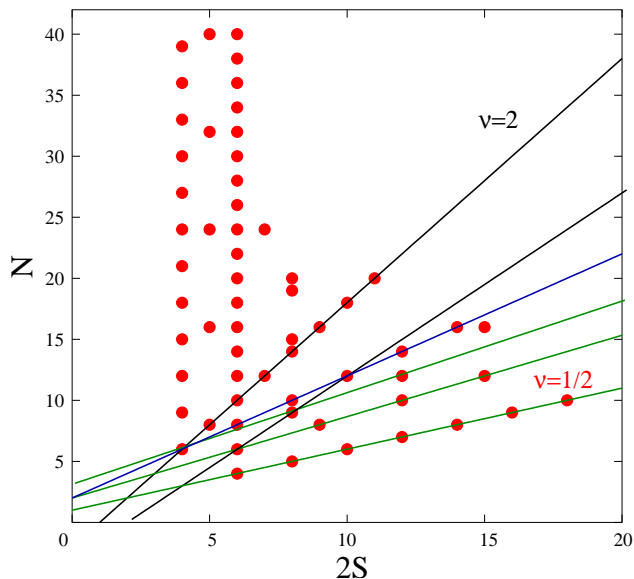}
\end{center}
\caption{Candidates for incompressible fluids in the
plane number of bosons vs flux quanta in the spherical geometry.
From bottom to top, we have the line $\nu =1/2$ then $\nu =2/3,
3/4$ then the $\nu =1$ states with the Pfaffian shift and finally
the hierarchical states at $\nu =3/2, 2$.}
\label{incomp}
\end{figure}

\section{The paired $\nu =1$ state}
In the Jain principal sequence of fractions there is an
accumulation point at $\nu =1$ at which the CFs feel zero net
flux. This is the Bose analog of the $\nu =1/2$ state for
fermions. We thus expect that either there is some kind of Fermi
sea of composite fermions as is the case for electrons with
Coulomb repulsion or it may also happen that this Fermi sea is
unstable towards pairing of CFs~\cite{Moore91,Greiter92}.
This latter case would be the
Bose analog of the paired state suggested to exist at the
enigmatic $\nu =5/2$ state of electrons. On the sphere, the
zero-flux on the CFs is realized for $2S=N-1$. When N is a square,
the CFs fill a closed shell and there are low-lying excitations
that have a characteristic structure~\cite{RezayiRead,Morf95}. One
expects that between the closed-shell states the angular momentum
of the ground state is determined by the second Hund's rule (max
L). While this law applies for fermions with Coulomb or hard-core
interaction, there are deviations in the case of bosons.
In addition there is no clear scaling to the infinite system for
the ground state energy. This can be taken as evidence that the
Fermi sea is unstable in the Bose case. A prototypical
wavefunction describing the pairing of the CFs is the so-called
Pfaffian~:
\begin{equation}\label{PfaffianEq}
    \Psi =\mathrm{Pf}\{\frac{1}{u_iv_j-u_jv_i}\}\prod_{i<j}(u_iv_j-u_jv_i)^q .
\end{equation}
The symbol Pf stands for the Pfaffian and $N$ should be even and we have defined
$u=\cos (\theta /2)\mathrm{e}^{i\varphi/2},
    v=\sin (\theta /2)\mathrm{e}^{-i\varphi/2}$.
Given an antisymmetric $N\times N$
matrix $A_{ij}$ it is defined by~:
\begin{equation}\label{PfDef}
\mathrm{Pf}\{A_{ij}\}=\sum_\sigma \epsilon_\sigma
A_{\sigma(1)\sigma(2)}\ldots A_{\sigma(N-1)\sigma(N)},
\end{equation}
where $\sigma$ are permutations of the index with $N$ values. For
$q=1$ the wavefunction Eq.(\ref{PfaffianEq}) describes bosons at
filling factor $\nu =1$ while for $q=2$ it describes a fermion state at
$\nu =1/2$. The filling factor is $\nu=1/q$ and this wavefunction will fit
on the sphere if we have~:
\begin{equation}\label{fluxPf}
    2S=q(N-1)-1.
\end{equation}
In the Bose case, there is a complete series of incompressible
states at the locations predicted by this equation. The gap
extrapolates to a nonzero value in the thermodynamic limit. In
addition, when one add or remove one flux quantum from the perfect
matching condition one observes the creation of two-particles
states contrary to the case of the (unpaired) Laughlin fluids.
This is a direct spectroscopic evidence for the paired nature of
the fluid at $\nu =1$.

\section{Incompressibility at $\nu > 1$}
For bosons it is possible to stay entirely into the LLL while
reaching filling factors greater than 1. The Jain sequence extends
naturally in this domain for $\nu =p/(p-1)$ and there are states
with spectra displaying the characteristics of incompressibility
on the lines predicted by Eq.(\ref{Hshift}). However they show
very irregular behavior as a function of the number of particles.
This novelty with respect to fermions lead to incompressible
states that may be the clustered Read-Rezayi states~\cite{ReadRezayi}. It has been
pointed out in the torus geometry that there are fractions for
$\nu =k/2$ where $k$ is an integer~\cite{Cooper01}. These states
have a ground state degeneracy which is exactly that expected for
the Read-Rezayi states and they have also high overlap with the
corresponding trial wavefunctions.The situation is not so clear in
the spherical geometry. For even higher filling factor there should be a
quantum phase transition towards the vortex lattice~\cite{Cooper01,Sinova02}
in the region $\nu \approx 6$.

\section{Rotating fermions}
It has been pointed out recently~\cite{Regnault04L} that a single
species of hyperfine fermions interacting through $p$-wave
scattering leads exactly to the hard-core model with only one
nonzero pseudopotential when projected onto the LLL. So the
celebrated Laughlin wavefunction for fermions is now the exact
ground state of the system for the filling 1/3. It is known that
the hard-core model displays FQHE properties that are close to
those of the Coulomb case. Here the principal sequence
has gaps whose order of magnitude is ruled by the combination~:
\begin{equation}\label{coupling}
    g_f= \sqrt{\frac{2}{\pi}}\frac{\hbar^2}{m}\frac{a_p^3}{\ell_z\ell^4}.
\end{equation}
Gaps were estimated for the principal sequence and if one is able
to enhance the $p$-wave scattering length up to values comparable
to the $s$-wave case then the FQHE for Fermi gases will have a
strength comparable to the Bose case. Concerning the fraction $\nu
=1/3$ one should note that it has charged excitations that are
gapped (quasielectrons) as well as gapless (quasiholes), similarly
to the Bose case. At exactly half-filling, there is a Fermi sea of
two-flux CFs. It appears on the sphere when the flux is given by
$2S=2(N-1)$. Using the model of free CFs one can extract the
effective mass of the CFs by considering the excitations above the
closed shell configurations that occur when N is a perfect square.
Finally we note that it has been recently suggested that the FQHE
should also appear in Fermi gases with dipolar
interactions~\cite{Baranov04}. If the interaction has a $1/r^3$ law
then it is intermediate between the pure hard-core forces that we
have discussed so far and the Coulomb interaction relevant to
electrons in semiconductor heterostructures. There is scattering
in all allowed partial waves (all pseudopotentials are nonzero)
hence the Laughlin state is no longer an exact eigenstate but it
will remain an excellent approximation. Quasiholes are now gapped
as is the case for the Coulomb interaction. For small filling
factor, it is likely that the Laughlin liquids will compete with
the Wigner crystal state. This is a very interesting situation
since this crystal, if realized, would be free of disorder and
hence amenable to detailed experimental and theoretical studies.

\section{Conclusions}

Trapped gases offer a so far theoretical opportunity to study the physics of
the FQHE in new conditions. The Bose gases display a new version of the
principal sequence of FQHE fractions and also paired states that have
so far no equivalent in the electronic world.
The elusive fractional statistics may be eventually observed experimentally~\cite{Paredes01}
The Fermi gases may be a perfect
incarnation of the hard-core model of the FQHE,
displaying notably the elusive Fermi sea of CFs at half-filling.
With dipolar fermions one may even study the competition with crystalline states at low
filling factor of the LLL.


We thank Yvan Castin and Jean Dalibard for numerous discussions.
Some of the numerical calculations have been performed thanks to a
computer time allocation of IDRIS-CNRS.


\end{document}